\documentstyle[12pt]{article}

\begin{document}
\title{Evidence for Excess Pions in Nuclei}
\author{S.V.Akulinichev\\
{\it Institute for Nuclear Research,}\\
{\it 60-th October Anniversary Prospect 7a, Moscow 117312, Russia.}
\\and\\
{\it Cyclotron Institute, Texas A\&M University, College Station}\\
{\it Texas 77843-3366, USA}
}
\maketitle
\begin{abstract}
We analize two reactions: the electroproduction at small $x$ and the
Drell-Yan production by protons on nuclei. We show that nuclear shadowing in
electroproduction is larger if the virtuality of intermediate
hadrons is properly taken into account.
In the Drell-Yan production, initial state interactions of fast partons
are included.
With this input, the excess pion contribution improves the
description of data for both reactions.

\end{abstract}
\newpage
1. Following the conventional picture of nuclei, the meson exchange currents
result in a certain amount of excess pions in nuclei.
This fundamental nuclear property was questioned by  some recent experiments.
The behaviour of nuclear structure functions
at low $x$\cite
{NMC} and a lack of nuclear cross section enhancement
for Drell-Yan (DY) production by protons
\cite{Alde} have led several authors (see, e.g., the Ref.\cite{BFS}) to the
conclusion that there are  no excess pions in nuclei.
The present analysis of these reactions is based on some new conjectures.
In electroproduction, it is shown that virtual hadrons
interact only inelastically with targets. As the result, the nuclear shadowing
can be significantly stronger than it has been believed so far.
In DY production by protons, as well as in other hadroproduction reactions
\cite{AZP}, initial state interactions of fast partons
are taken into account.
With this input, the data for both reactions are much better described with the
excess
pion contribution included.

2. In the impulse approximation, the quark (antiquark) distribution in nuclei
can be represented as a convolution of nucleon and excess pion quark
distributions,
\begin{equation}
q^{A}(x)=\int \frac{dy}{y} f^{A}_{N}(y)q^{N}(\frac{x}{y}) + \int \frac{dy}{y}
f^{A}_{\pi}(y)q^{\pi}(\frac{x}{y}),
\end{equation}
where $f^{A}_{N}(y)$ and $f^{A}_{\pi}(y)$ are the nucleon and excess pion
distribution functions in nuclei. The flavour indices and the $Q^{2}-$
dependence are implied  here and hereafter.
Note that we denote by $f^{A}_{\pi}(y)$ the renormalized
excess pion distribution which preserves the total probability
conservation\cite{SS}.
Therefore the functions $f^{A}_{N}(y)$ and $f^{A}_{\pi}(y)$ are not
independent but are constrained by the momentum sum rule
\begin{equation}
\int dy y f^{A}_{N}(y) +\int dy y f^{A}_{\pi}(y)\equiv <y^{A}_{N}>
+<y_{\pi}^{A}>=1.
\end{equation}
One way to find $q^{A}(x)$ is to calculate first
$f^{A}_{\pi}(y)$ using the elementary $\pi N$ interaction and
then to determine $f^{A}_{N}(y)$ from (2), adopting some model for the
nucleon momentum distribution. In earlier works of this kind\cite{ET}
the unrenormalized pion distribution was used for
$f^{A}_{\pi}(y)$, what led to an overestimation of
the pionic contribution to $q^{A}(x)$.
This was corrected in more recent papers\cite{Bub}, but there is a
significant model dependence of calculations,
which start from the elementary $\pi N$ interaction.

Another approach was suggested in Ref.\cite{AKV}. Since $f^{A}_{N}(y)$ is
an invariant function ($y$ is the ratio of two scalar products),
it can be calculated in any frame. In the nucleus rest frame, $f^{A}_{N}(y)$
is determined by the energy and momentum distribution
of nucleons, which can be measured in independent experiments\cite{Fur}.
In this case the magnitude of the EMC-effect is proportional to
$(1-<y^{A}_{N}>)/m_{N}=\epsilon/m_{N}$, where $\epsilon$ is the average nucleon
separation energy. For heavy nuclei the data give $\epsilon\geq$40 MeV.
The A-dependence of $\epsilon$ was discussed in Ref.\cite{SV}.
When $f^{A}_{N}(y)$ is determined,
the first moment of the excess pion distribution is fixed
by (2) and a detailed form of $f^{A}_{\pi}(y)$ can be found by using some
realistic
form of pion momentum distribution, e.g. the one from Ref.\cite{FPW}.
Here we calculate $f^{A}_{\pi}(y)$ with this method.

3. It was shown\cite{AKV} that the main part of the EMC-effect
can be explained by the binding correction of nuclear structure functions.
We here focus on the low-$x$ region and show
that the nuclear shadowing
and antishadowing can be explained provided (a) the virtuality of
intermediate mesons is taken into account while considering their
multiple interactions in nuclei and (b) the contribution of excess
pions is added. We use the generalized vector meson dominance (GVMD) model, but
the above conclusion is valid for some other hadron dominance models,
adopting the Glauber theory\cite{Glaub}
to describe multiple interactions of virtual hadrons.
In the GVMD model, the structure function is given by (see, e.g., the Ref.
\cite{GVMD})
\begin{equation}
F_{2}^{GVMD}(x)=\frac{Q^{2}}{\pi}\int^{\infty}_{0}d\mu^{2}\frac{\mu^{2}\Pi(\mu^
{2})}{(\mu^{2}+Q^{2})^{2}}\sigma_{tot}(\mu^{2},s),
\end{equation}
where $\Pi(\mu^{2})$ is the spectral function of mesonic states with the mass
$\mu$, $\sigma_{tot}$ is the total cross section of meson
interaction with a target.
Total cross sections are connected to
forward scattering amplitudes by the optical theorem,
$\sigma_{tot}=Im[f(s,t=0)]/s$ (we use the normalization
$<p|p'>=(2\pi)^{3}2p_{0}\delta(\vec{p}-\vec{p'}))$. In some other
models\cite{SF},
the parton-target amplitudes have been used instead.
In the independent particle model, the Glauber result,
adopted by many authors, is given by
\begin{equation}
f^{A}(s,t)=-2si\sum_{n=1}^{A}C^{A}_{n}\int d\vec{b}(\frac{if^{N}(s,t)
T(\vec{b})}{2sA})^{n},
\end{equation}
where $T(\vec{b})$ is the optical thickness of the nucleus and $\vec{b}$
is the impact parameter. For $Im[f^{N}]/Re[f^{N}]\gg 1$ and $A\gg 1$,
(4) can be rewritten as
\begin{equation}
\sigma^{A}_{tot}=2\int d\vec{b}(1-e^{-\sigma^{N}_{tot}T(\vec{b})/2}).
\end{equation}
We now show that (4) and (5) cannot be used for virtual hadrons participating
in electroproduction.

In Glauber theory it is assumed that for fast projectiles the
overall modification of the wave inside the nucleus is the factor
\cite{Glaub,Yenn}
\begin{equation}
1-\Gamma_{A}(\vec{b})=\prod^{A}_{i=1}[1-\Gamma_{N}(\vec{b}-\vec{s}_{i})],
\end{equation}
where $\Gamma_{N}(\vec{b}-\vec{s}_{i})$ is the profile function of a nucleon
with the transverse coordinate $\vec{s}_{i}$. This equation is also valid for
virtual projectiles if their propagation length $\lambda$
satisfies the condition $\lambda\gg r_{A}$, where $r_{A}$ is the nucleus
size. A correction of (6) in the case $\lambda\sim r_{A}$ has been considered
in Ref.\cite{Grib} and will be taken into account below.
The equation (4)  follows
from (6) if the elastic scattering amplitudes are expressed in terms of profile
functions as\cite{Glaub,Yenn}
\begin{equation}
f^{N,A}(s,t)=\frac{i|\vec{k}|}{2\pi}\int d^{2}\vec{b} e^{-i\vec{k}\vec{b}}
\Gamma_{N,A}(\vec{b}),
\end{equation}
where $\vec{k}$ is the projectile three-momentum in the lab. frame.
The last equation describes the Fraunhofer diffraction of waves\cite{Landau},
if the diffraction is observed
at large distances $z$ from the target.  The condition for the
Fraunhofer diffraction  (and for the validity of (7)) is\cite{Landau}
\begin{equation}
z\gg L,\:L=2kr^{2}_{t},
\end{equation}
where $L$ is the healing length of shadow\cite{Yenn} ($L$ characterizes
the duration of the elastic scattering process) and $r_{t}$ is the target
transverse size. At $z\sim L$ the Frenel diffraction, which is weaker than
the Fraunhover diffraction, takes place instead
and at $z\ll L$ there is no diffraction\cite{Landau}. But for virtual
projectiles the accessible longitudinal distances are only of the order
of $\lambda$.  Therefore the standard derivation of (4) doesn't hold if
\cite{Akul}
\begin{equation}
\lambda \ll L.
\end{equation}
In that case, a projectile vanishes before the diffraction takes place.
Let us obtain the same result using a more formal language.

For stable projectiles, the Glauber result for elastic and inelastic
cross sections reads
\begin{equation}
\sigma_{el}=\int d^{2}\vec{b}|\Gamma(\vec{b})|^{2},\:\sigma_{inel}=
\int d^{2}\vec{b}(1-|1-\Gamma(\vec{b})|^{2}).
\end{equation}
As it follows from this equation,
both $\sigma_{el}$ and $\sigma_{inel}$ are non-vanishing if
$\Gamma(\vec{b})\neq 0$.
However, this is not the case for
virtual projectiles and we now show that $\sigma_{el}$ vanishes
under the condition (9), but $\sigma_{inel}$ does not.
The total interaction probability is given by the imaginary part
of the forward projectile-target amplitude $f(s,t=0)\equiv<k,P|\hat{T}|k,P>$
where $k,P$ are projectile and target momenta and the operator $\hat{T}$
describes all projectile-target interactions. The imaginary part of the
amplitude can be represented as the sum of elastic and inelastic channels,
\[ Im<k,P|\hat{T}|k,P>=\frac{(2\pi)^{4}}{2}\sum_{n}\delta^{4}(k_{n}+P_{n}-
k-P)<k,P|\hat{T}^{\dagger}|k_{n},P_{n}><k_{n},P_{n}|\hat{T}|k,P> \]
\begin{equation}
+\frac{(2\pi)^{4}}{2}\sum_{i}\delta^{4}(P_{i}-k-P)<k,P|\hat{T}^{\dagger}|i>
<i|\hat{T}|k,P>,
\end{equation}
where $|k_{n},P_{n}>$ are all possible states of initial particles and
$|i>$ are all other states, $|i>\neq |k_{n},P_{n}>$.
The intermediate states $|k_{n},P_{n}>$ and $|i>$ contribute to $Im[f(s,t=0)]$
if they contain only free particles. In fact, to take the
imaginary part of a diagram is equivalent to cut this diagram and to put the
cut lines on the mass shell.
In the case of virtual hadron-nucleus scattering, $P^{2}=m_{A}^{2}$ but
$k^{2}\neq\mu^{2}$, where $\mu$ is the hadron mass. It is easy to verify
that the first sum from (11) vanishes when $P^{2}_{n}=m_{A}^{2},
k_{n}^{2}=\mu^{2}$. A hadron with the momentum $k_{n}$
must be off-mass-shell (the opposite
case of an off-mass-shell recoil nucleus is irrelevant for the process under
consideration). Due to the uncertainty principle, a hadron with
the off-mass-shell
momentum $k_{n}$ can be considered as a free particle during the time
$\lambda\sim 1/(k_{n,0}-(\vec{k}^{2}_{n}+\mu^{2})^{1/2})$. For fast particles,
$\lambda$ is their propagation length. In our case $\lambda$ is similar for
hadrons carrying $k_{n}$ and $k$. Thus, for virtual projectiles the
contribution of the elastic channel, represented by the first sum from
(11), is non-vanishing during the time of the order of
$\lambda$.  If $L$ is the duration of the scattering process, then the elastic
cross section vanishes in the case (9), in agreement with the above
euristic argumentation. In this case the only contribution to $Im[f(s,t=0)]$
is due to inelastic channel because the states $|i>$ contain new particles,
which can be on the mass shell. When (9) is fulfilled,
we obtain
\begin{equation}
Im[f^{A}(s,t=0)]=s \sigma^{A}_{inel},\:\sigma_{el}^{A}\approx 0.
\end{equation}
In electroproduction on nuclei, $\lambda\sim 1/xm_{N}$ and $L_{A}=2q_{0}
r^{2}_{A}$ where $q$ is the virtual photon momentum. In the shadowing region
$L_{A}\geq 100\lambda$ and we have to use $\sigma_{inel}^{A}$ instead of
$\sigma_{tot}^{A}$ to calculate $F_{2}^{A}(x)$. In most cases, the
condition (9) is also valid for separate nucleons and the free
nucleon structure function determines $\sigma_{inel}^{N}$  rather than
$\sigma_{tot}^{N}$ for virtual hadrons.  Note that the condition
for the Fraunhofer diffraction never holds in electroproduction.

In the independent particle approximation and for $A\gg 1$, the Glauber result
for the nuclear inelastic cross section is given by
\begin{equation}
\sigma_{inel}^{A}=\int d\vec{b}(1-e^{-\sigma^{N}_{inel}T(\vec{b})}).
\end{equation}
In contrast to (5), the last equation is correct for any
value of $Im[f^{N}]/Re[f^{N}]$.
In the exponent in (13) there is no $1/2$-factor, which
is present in (5) due to diffraction effects.
If $\lambda/r_{A}$ is finite, for the square well
form of nuclear density the equation (13) can be rewritten as
\begin{equation}
\sigma_{inel}^{A}=2\pi\rho\sigma_{inel}^{N}\int^{r_{A}}_{0}dy\:y\int^{2y}_{0}
dz exp[-\rho\sigma_{inel}^{N}z\: exp(-z/\lambda)],
\end{equation}
where $\rho$ is the nuclear density. The
exponent inside the exponent simulates the $x$-dependence of nuclear shadowing
due to the finiteness of $\lambda/r_{A}$.
To calculate nuclear structure functions $F^{A,GVMD}_{2}(x)$
at low $x$, we have used (3) with $\sigma_{tot}$ equal to $\sigma_{inel}
^{A}$ from (14).

Following the data for high energy hadron-nucleon interactions,
we assume that for virtual mesons $\sigma_{inel}^{N}\approx
5/6\sigma_{tot}^{N}$ as well as for free mesons. The final results are not
sensitive to small variations of $\sigma_{inel}^{N}$.
We have used  the standard set of parameters for low lying mesons.
$\Pi(\mu^{2})$ for the meson continuum
was determined from the data for the $e^{+}e^{-}$-annihilation following
the prescription of Ref.\cite{GVMD}.
The cross section for the continuum was taken in a form
$\sigma^{N}_{inel}=18 mb\: GeV^{2}/\mu^{2}$ in order to reproduce the
experimental nucleon structure function.
For every $x$-bin, the average $Q^{2}$ is the same as in Ref.\cite{NMC}.
At $x>0.175$, when the contribution of virtual hadrons becomes negligible,
we have calculated $F_{2}^{A,EMC}(x)$ using the results of
Ref.\cite{AKV} with $\epsilon$=35 MeV, with the Fermi momentum $p_{F}$=230
MeV/c
and with the flux factors\cite{SFA} included.
Our result for the structure function ratio can be represented as
\begin{equation}
R= (F_{2}^{A}(x) + \Delta F_{2}^{A,\pi}(x))/A F_{2}^{N}(x),
\end{equation}
where $F_{2}^{A}(x)=\theta(0.175-x)F^{A,GVMD}_{2}(x)+\theta(x-0.175)
F^{A,EMC}_{2}(x)$. The excess pion contribution $\Delta F_{2}^{A,\pi}
(x)$ was calculated following the preceeding section with $\epsilon$=
35 MeV and with the pion three-momentum distribution from Ref.\cite{FPW}.
In some previous calculations, on-mass-shell pions with $E_{\pi}=\sqrt{\vec{p}
_{\pi}^{2}+m_{\pi}^{2}}$ \cite{AKV} or static pions with $E_{\pi}=0$ \cite{ET}
have been used. Here we take the intermediate value $E_{\pi}=\sqrt{
\vec{p}^{2}_{\pi}+m_{\pi}^{2}}/2$. In this case $R$ deviates by less than
2\% from the ratio calculated for the two extreme values of $E_{\pi}$.
For the number of excess pions per nucleon we obtained $n_{\pi}\approx 0.07$.
The elementary structure functions were taken from Ref.\cite{free}.

The resulting $R$ is shown by the solid line on Fig.1.
The nucleon contribution, represented by the first term from
(15), is shown by the dashed line. The earlier GVMD model result\cite{GVMD}
for the nuclear shadowing effect is shown by the dotted line.
In our method
the nuclear shadowing is significantly larger than in previous calculations,
because the shadowing is more effective in the inelastic channel.
The agreement with the data is restored by the virtual pion contribution.
In this case the antishadowing at $x\sim 0.1$ is also
explained. Though we obtained a good description of the data,
we understand that the GVMD model may be too simplified
and can hardly provide a
correct $Q^{2}$-dependence of nuclear shadowing.
But the result of this section is more general then the GVMD model:
following our conjecture, the nuclear shadowing will be  larger
in  other hadron dominance models based on Glauber theory.
The agreement with data, obtained in some papers\cite{GVMD,SF},
can be recovered if the excess pion contribution will be included.

4. The DY production by protons probes the antiquark content of a target
and for the cross sections of this reaction it was usually accepted
\begin{equation}
R_{DY}\equiv \sigma_{DY}^{A}/A\sigma_{DY}^{N}
\approx\overline{q}^{A}(x)/A\overline{q}^{N}(x).
\end{equation}
The measured $R_{DY}$\cite{Alde} is close to 1 (see Fig.2). This would mean
that there is no excess pions if (16) is correct.
In Ref.\cite{AZP} it is shown that there is no reason to neglect
absorptive initial state (IS) interactions of fast partons in nuclei.
The conclusion of that paper is that the color transparency forecasts
restrict the strength of final state interactions
but do not rule out IS interactions
in hadroproduction. With the absorptive IS interactions included, for $R_{DY}$
we obtain\cite{AZP}
\begin{equation}
R_{DY}=F \overline{q}^{A}(x)/A\overline{q}^{N}(x).
\end{equation}
The factor $F$ describes the projectile parton flux attenuation in nuclei,
\begin{equation}
F=\frac{1}{A\sigma^{q}_{abs}}\int d^{2}\vec{b}(1-e^{-
\sigma^{q}_{abs}T(\vec{b})}),
\end{equation}
where $\sigma^{q}_{abs}$ is the cross section of absorptive interactions of
fast initial partons with nucleons.
We assume that $\sigma_{abs}^{q}$ is the same for quarks and antiquarks.
In this case the parameter $\sigma^{q}_{abs}$ can be found
from the DY productions by pions, which is not affected by the excess pion
contribution.
The analysis of Ref.\cite{AZP} gives $\sigma_{abs}^{q}\approx$ 2mb and we used
this value to calculate $R_{DY}$ with (17) and (18).
In our calculations, the nuclear antiquark density
$\overline{q}^{A}(x)$ includes the excess pion contribution by analogy to (1).
This contribution was calculated as described in preceeding sections.
We considered only u-quarks and antiquarks with the parametrizations
from Ref.\cite{free}.
The binding correction for $\epsilon$=35 MeV was introduced as described above.
The ratio $R_{DY}$ calculated without the excess pion contribution (the
lower dashed line on Fig.2) is significantly below the data points.
The full result for $R_{DY}$ calculated with $E_{\pi}=\sqrt{\vec{p}^{2}_{\pi}+
m_{\pi}^{2}}/2$ is shown by the solid line. The same for $E_{\pi}=0$ is
shown by the upper dashed line.
As in the case of electroproduction, the pionic contribution improves the
agreement with data. The bulk of the pion
contribution is fixed by the sum rule (2) but the detailed form of resulting
curves depends on the choice of the pion distribution.

We conclude that both electro- and hadroproduction data, as well as the data
for $K^{+}-A$ scattering\cite{Akaon}, can be
explained with the excess pion contribution included.

I would like to thank the Cyclotron Institute of Texas A\& M University,
where I began this work, for the warm hospitality.

This work was funded in part by the Russian Foundation of Fundamental Research
(conract No 93-02-14381).

\newpage
{\bf Figure captions}

Fig.1. The ratio $F_{2}^{A}(x)/A F_{2}^{N}(x)$ calculated
without (dashed line) and with (solid line) excess pions. The dotted line
is the earlier result of the GVMD model\cite{GVMD}. The data are from Ref.\cite
{NMC}.

Fig.2. The ratio $\sigma_{DY}^{A}/A\sigma_{DY}^{N}$ for the Drell-Yan
production by protons calculated without excess pions (lower dashed line)
and with two different forms of excess pion distribution (solid and upper
dashed
lines). The earlier estimation of this ratio with excess pions included
(dotted line) and the data points are from Ref.\cite{Alde}.
\end{document}